\documentclass[aoas,MSNbibl,nameyear,dvips]{arximspdf}
\usepackage{graphicx}
\doi{10.1214/12-AOAS604} %kopijuoti is PTS
\volume{7}
\issue{2}
\pubyear{2013}
\firstpage{883}
\lastpage{903}

\begin{document}
\begin{frontmatter}

\title{Hierarchical Bayesian analysis of somatic mutation data in
cancer\thanksref{T1}}
\runtitle{Bayesian analysis of somatic mutations}

\thankstext{T1}{Supported by the National Science Foundation [award 0342111].}

\begin{aug}
\author[A]{\fnms{Jie} \snm{Ding}\ead[label=e1]{jding@jimmy.harvard.edu}},
\author[A]{\fnms{Lorenzo} \snm{Trippa}\ead[label=e2]{ltrippa@jimmy.harvard.edu}},
\author[B]{\fnms{Xiaogang} \snm{Zhong}\ead[label=e3]{xz248@georgetown.edu}}\\
\and
\author[A]{\fnms{Giovanni} \snm{Parmigiani}\corref{}\ead[label=e4]{gp@jimmy.harvard.edu}}
\runauthor{Ding, Trippa, Zhong and Parmigiani}
\affiliation{Dana-Farber Cancer Institute and Harvard School of Public
Health, Dana-Farber Cancer Institute and Harvard School of Public
Health, Georgetown University, and Dana-Farber Cancer
Institute and
Harvard School of Public
Health}
\address[A]{J. Ding\\
L. Trippa\\
G. Parmigiani\\
Department of Biostatistics \\
\quad and Computational Biology\\
Dana-Farber Cancer Institute\\
Boston, Massachusetts 02215\\
USA\\
and\\
Department of Biostatistics\\
Harvard University\\
Boston, Massachusetts 02115\\
USA\\
\printead{e1}\\
\hphantom{E-mail: }\printead*{e2}\\
\hphantom{E-mail: }\printead*{e4}}
\address[B]{X. Zhong\\
Department of Biostatistics\\
\quad Bioinformatics and Biomathematics\\
Georgetown University\\
Washington, DC 20057\\
USA\\
\printead{e3}} %adresu isvedimo komanda gale!
\end{aug}

% HISTORY:
\received{\smonth{3} \syear{2012}}
\revised{\smonth{9} \syear{2012}}

% ABSTRACT
%
\begin{abstract}
Identifying genes underlying cancer development is critical to
cancer \mbox{biology} and has important implications across prevention,
diagnosis and treatment. Cancer sequencing studies aim at
discovering genes with high frequencies of somatic mutations in
specific types of cancer, as these genes are potential driving
factors (drivers) for cancer development. We introduce a
hierarchical Bayesian methodology to estimate gene-specific mutation
rates and driver probabilities from somatic mutation data
and to shed light on the overall proportion of drivers among
sequenced genes. Our methodology applies to different experimental
designs used in practice, including one-stage, two-stage and
candidate gene designs. Also, sample sizes are typically small
relative to the rarity of individual mutations. Via a shrinkage
method borrowing strength from the whole genome in assessing
individual genes, we reinforce inference and address the selection
effects induced by multistage designs. Our simulation studies show
that the posterior driver probabilities provide a nearly unbiased
false discovery rate estimate. We apply our methods to
pancreatic and breast cancer data, contrast our results to previous
estimates and provide estimated proportions of drivers for these two
types of cancer.
\end{abstract}

% KEYWORDS
% Pirmas kwd is didziosios raides
%

\end{frontmatter}

%s1 #&#
\section{Introduction}
\label{sec1}

We introduce a semiparametric hierarchical Bayesian model for the
analysis of somatic mutations in cancer. Our study is motivated by
experiments sequencing comprehensive libraries of coding genes in
tumors and matching normal samples
[Cancer Genome Atlas Research Network (\citeyear{CancerGenomeAtlasResearchNetwork2008,CancerGenomeAtlasResearchNetwork2011}),
\citet{Sjoblom2006,Greenman2007,Wood2007,Jones2008,Parson2008,Kan2010}].
A main goal of these studies has been to provide lists of candidate
cancer genes, for which evidence of a role in driving carcinogenesis
emerged from the the presence of somatically acquired differences
between tumor and normal genomes. These driver genes need to be
distinguished from so-called passenger genes, which present somatic
mutations in cancer even though these mutations are not directly
related with the tumor genesis. Statistical tools for this task have
been based on hypotheses testing theory and, in particular, on methods
for controlling the false discovery rates (FDR) of reported gene lists
[\citet
{Greenman2006,Wood2007,Getz2007,Parmigiani2009,Trippa2011}]. Our
goal here is to complement this approach with methodology for
deriving the probability that a gene contributes to carcinogenesis.
There are four important reasons for this: to handle multistage
designs; to remedy the severe FDR overestimation resulting from
one-gene-at-a-time analyses; to improve ranking and selection of genes
for subsequent analyses; and to address estimation of the total number
of cancer drivers.

First, the rarity of mutations and the cost of sequencing comprehensive
lists of genes have motivated the use of multistage designs, to balance
between resource use and power in detecting cancer genes
[\citet{Kraft2006,Skol2006,Wang2006,Sjoblom2006,Parmigiani2009}]. In
these studies, genes are selected for later stages based on results of
earlier stages as well as a host of other biological considerations,
including membership in key pathways, potential for drug targeting,
reliability of sequencing and
findings of previous sequencing studies. \citet{Kan2010}, for
instance, discussed the analysis of 1507 genes selected in part on the
basis of previously published results. Methods based on $p$-values do
not include prediction of the final findings at completion of the
first stage. This limit, in multi-stage problems, compounds with
conceptual challenges when biological judgment is used to refine
lists of candidates that are moved along to the last stages of the
study. Also, multiple hypothesis testing methods are not designed for
optimally selecting genes for subsequent stages, while Bayesian analysis
allows one to obtain the probabilities (i) that a gene is a driver
and (ii) that it will be validated in subsequent stages.
Posterior driver probabilities provide two unique advantages. Prior to
a new study or stage with a pre-specified hypothetical sample
size, they allow, unlike $p$-values, to assess the probability, for each
gene, of finding a number of mutations that would provide evidence of
an abnormal mutation rate. After a study, they are applicable for
summarizing the study findings, irrespective of the selection criteria
used to move genes through stages.

Second, the standard inferential approach for mutation analysis is to
compute false discovery rates based on standard multiple testing
correction, following one-gene-at-a-time analyses such as
likelihood ratio tests. This approach can lead to a severe
overestimation of the FDR.\eject

Third, an important goal of somatic mutation analysis is to determine
genes' mutation rates. In a typical genome-wide study, sample sizes are
small relative to the rarity of individual mutations. For example, we
expect to observe no mutations for most of the genes, though estimating
a population-level mutation rate of zero would be biologically
implausible. Also, in multi-stage designs, it is important to account
for possible biases arising from selecting genes with high mutation
frequencies in early stages. Both issues can be addressed using a
model-based approach for estimating individual genes' mutation rates
by ``borrowing strength'' from the entire set of mutations across
the genome [\citet{Efron1973}]. Shrinkage affects posterior driver
probabilities and mutation rates estimates, as genes for which less
information is available are pulled more strongly toward the
genome-wide average.

Fourth, the change of landscape resulting from early cancer genome
projects has posed the question of the proportion of driver genes across
the genome. \citet{Wood2007} had pointed at this question and
proposed conservative estimators applied for FDR control with
empirical Bayes testing procedures. Our methodology is designed to
also provide an estimate of this proportion with the associated
statement of uncertainty.

The organization of the remaining sections is as follows. Section \ref{sec2}
gives a general description of cancer somatic mutation data. Section
\ref{sec3}
describes our Bayesian hierarchical model. Section \ref{sec4} shows the
results of simulated experiments designed to assess the improvement
provided by our approach over standard alternatives. Section \ref{sec5}
presents a re-analysis of two published sequencing studies. Finally,
Section \ref{sec6} provides additional discussion about our method and results.

%s2 #&#
\section{Cancer somatic mutation data}
\label{sec2}
We consider studies providing a collection of somatic mutations from
genome-wide exome sequencing of samples of a specific tumor
type. Somatic mutations can be detected by comparing DNA sequences of
tumor samples to those of their matching normal samples. Each mutation
is labeled as one of a set of possible mutation types, as in the
example of Table \ref{tabmuttype}. Mutations of different types are
observed to
%
%t1 #&#
\begin{table}
\tablewidth=139pt
\caption{24 point mutation types}
\label{tabmuttype}
\begin{tabular*}{\tablewidth}{@{\extracolsep{\fill}}lcccc@{}}
\hline
\textbf{Mutated from}&\multicolumn{4}{c@{}}{\textbf{Mutated to}}\\
\hline
C in CpG&A&--&G&T\\
G in CpG&A&C&--&T\\
G in GpA&A&C&--&T\\
C in TpC&A&--&G&T\\
A&--&C&G&T\\
Other C&A&--&G&T\\
Other G&A&C&--&T\\
T&A&C&G&--\\
\hline
\end{tabular*}
\end{table}
have varying overall frequencies in tumor samples. Different
definitions of mutation types may be used to suit different data
structures or
different biological questions. In this paper, as in \citet{Wood2007}
and \citet{Jones2008}, each mutation is classified either as a small
insertion/deletion or as one of 24 types of single nucleotide changes,
defined in Table \ref{tabmuttype}. For each gene, mutation type and
sample, it is
important to consider the mutation count as well as the number of
nucleotides at risk for that type of mutation, heretofore called the
coverage. The coverage for a gene may be smaller than the total base
count because not all bases may be reliably sequenced.

We analyze data generated in two previous studies. The first
[\citet{Jones2008}] includes 24 tumors with matching normal
tissues from
patients with pancreatic malignancies. The study sequenced 20,671 genes
and found 1163 nonsynonymous somatic mutations harbored in 1007 genes.
These mutations were categorized by gene, mutation type and sample. The
second study [\citet{Wood2007}] considered breast cancer, and
adopted a
two-stage design with 11 samples in the discovery stage and 24 samples
in the subsequent validation stage. During the discovery stage, 18,190
genes were sequenced and 1112 nonsynonymous mutations were identified
in 1026 genes. During the validation stage, these 1026 genes were
sequenced in the additional 24 tumors, and 190 nonsynonymous mutations
were identified in 154 genes. Mutations were categorized by gene,
mutation types and stage. The data,
at the gene level, include two mutation counts, one for each stage.
An advantage of performing Bayesian analyses of these data sets
is that both the probability model and the computational procedures can be
straightforwardly adapted to these designs, as well as other
multi-stage designs.

%s3 #&#
\section{Model}
\label{sec3}

Somatic mutation counts are modeled using a Bayesian multilevel
semi-parametric model. At the data level, the observed count of
somatic mutations of type $m$ in gene $g$ and sample $k$, indicated by
$X_{\mathit{gmk}}$, has distribution
%
%e1 #&#
\begin{eqnarray}
X_{\mathit{gmk}} &\sim& \operatorname{Poisson}(\lambda_{\mathit{gmk}}T_{\mathit{gmk}}),
\nonumber\\[-8pt]\\[-8pt]
&&\eqntext{g=1,\ldots,G; m=1,\ldots,M; k=1,\ldots,K,}
\end{eqnarray}
where $\lambda_{\mathit{gmk}}$ is the unknown mutation
rate and $T_{\mathit{gmk}}$ is the observed coverage for the corresponding
gene, mutation type and sample, that is, the number of successfully
sequenced bases in gene $g$ and sample $k$, that are susceptible to
a mutation of type $m$.
The term ``coverage,'' in the next-generation sequencing
literature, has a different interpretation. Here we use it
consistently with earlier studies using Sanger
sequencing technology [e.g., \citet{Wood2007}].

The binomial and multinomial distributions are often used for mutation
counts in somatic mutation analysis [\citet{Greenman2006}]. Here
we use a
Poisson distribution because it is a good approximation of both those
distributions when the mutation rates are small and
because it simplifies the calculation of the posterior distributions.
Our model assumes that mutations within a single gene and among
different genes occur independently of each other conditional on
mutation rates.

At the mutation rate level, we use a multiplicative random effects model
%
%e2 #&#
\begin{equation}
\lambda_{\mathit{gmk}} = \lambda_g \alpha_m
\beta_k,
\end{equation}
which includes a gene specific mutation rate $\lambda_g$, a mutation
type effect $\alpha_m$ and a sample effect $\beta_k$. The three
multiplicative components have the following interpretation: the
$\lambda_g$'s
allow to assign each gene its own mutation rate; the $\alpha_m$'s allow
%for the mutation rates to vary with the mutation type
the rates to vary across mutation types; the $\beta_k$'s
allow different samples to have different mutation rates, a feature
observed in most data sets. We set $\prod_{m=1}^M \alpha_m = 1$ and
$\prod_{k=1}^K\beta_k = 1$ to make the model identifiable.

We propose and compare two complementary approaches, one for estimating
gene-specific mutation rates and one for estimating gene-specific
driver probabilities. One of the main differences is that the first
approach does not require a reliable estimate of the passenger mutation
rate while the second does. The assumption of known passenger rates has
also been used in the previous literature [e.g.,
\citet{Jones2008,CancerGenomeAtlasResearchNetwork2008}] for identifying
driver genes with FDR methods.

To complete the multilevel model, we specify a distribution for
mutation rates across genes. We treat this distribution as unknown
and estimate it from the data with minimal distributional
assumptions. Early cancer genome
studies, for example, \citet{Wood2007}, have shown the existence
of small
subgroups of driver genes, the so-called ``mountains,'' with rates of
mutations over 100-fold higher than the assumed passenger rates. In
contrast, most of the likely drivers are found to harbor mutations
only in small proportions of samples and hence are called ``hills.''
This motivates the use of nonparametric modeling to mitigate the
overall influence of mountains on the inference.
We use a Dirichlet Process [\citet{Ferguson1973}] for
the unknown distribution of the mutation rates across the genome:
%
%e3 #&#
\begin{eqnarray}
F &\sim&\mbox{Dirichlet Process} \bigl(a,\operatorname{Exponential}(\gamma)
\bigr),
\nonumber\\[-8pt]\\[-8pt]
\lambda_g | F &\stackrel{\mathrm{i.i.d.}} {\sim}& F,
\nonumber
\end{eqnarray}
where $a$ is the so-called concentration parameter and $\gamma$
controls the mean of the random distribution $F$, chosen to be
exponential. The nonparametric Dirichlet prior is flexible
and has proven useful in several applications
modeling random effects distribution, as done here. See
\citet{Dunson2010} for an extensive overview.\eject

We can now consider the second case, in which the main interest
is to derive driver probabilities at the gene level.
Here we make the additional assumption that for all passenger genes,
$\lambda_g = \lambda_0$, a known mutation
rate. If this assumption holds, the driver genes
can be defined statistically as those with mutation
rates greater than $\lambda_0$, because any gene whose mutations have
the ability to provide a fitness advantage to cancer cells will occur
in cancer at adjusted rates higher than $\lambda_0$ when a large
enough population is considered. The word adjusted here refers to the
fact that, because of different coverage and nucleotide composition,
different passengers may still exhibit different mutation rates per
nucleotide even though the baseline mutation rate $\lambda_0$ is
common to all.

To derive driver probabilities,
we slightly modify the model above and include an additional
hierarchical level. We use binary variables $\delta_g$, one for each
gene, for distinguishing the drivers ($\delta_g=1$) from the
passengers ($\delta_g=0$). The $\lambda_g$ is now
%
%e4 #&#
\begin{equation}
\lambda_g = \mathrm{I}(\delta_g=0)\lambda_0
+ \mathrm{I}(\delta_g=1) \bigl(\lambda_g^d +
\lambda_0\bigr),
\end{equation}
where $\lambda_g^d$ is the difference between the mutation rate of a
putative driver $\lambda_g$ and the pre-specified underlying passengers rate
$\lambda_0$. Since $\delta_g$ is also unknown, a natural choice for
modeling the binary variables is the conjugate Beta-Bernoulli prior
%
%e5 #&#
\begin{eqnarray}
\pi&\sim&\operatorname{Beta}(a_{\pi},b_{\pi}),
\nonumber\\[-8pt]\\[-8pt]
\delta_g | \pi&\stackrel{\mathrm{i.i.d.}} {\sim}&\operatorname{Bernoulli}(\pi),
\nonumber
\end{eqnarray}
where $\pi$ is the unknown overall proportion of drivers among all genes.
We use a Dirichelet prior for the latent $\lambda_g^d$'s:
%
%e6 #&#
\begin{eqnarray}
F &\sim&\mbox{Dirichlet Process}\bigl(a,\operatorname{Exponential}(\gamma
)\bigr),
\nonumber\\[-8pt]\\[-8pt]
\lambda_g^d | F &\stackrel{\mathrm{i.i.d.}} {\sim}& F.
\nonumber
\end{eqnarray}

Diffuse flat prior densities are used for random vectors
$(\alpha_1,\ldots,\alpha_M)$ and $(\beta_1,\ldots,\beta_K)$.
We also use a Gamma hyper-prior for $\gamma$. The value
of $a$ in the Dirichlet process is set to 1. In simulations, we
considered several values of $a$ and performed sensitivity analyses.
We observed negligible variations in our results across prior
parameterizations.

The posterior distributions of parameters from the
hierarchical Bayesian models are estimated using a Markov Chain Monte
Carlo algorithm. See the supplementary material [\citet{Ding2012}]
for details.

%s4 #&#
\section{Simulation study}
\label{sec4}

%s4.1 #&#
\subsection{Scenarios}

We used simulations for validating our Bayesian procedure. Simulation
scenarios have strong similarities with the pancreatic study in\vadjust{\goodbreak}
\citet{Jones2008}. We used the same set of genes and their corresponding
coverage. We set the passenger mutation rate to $\lambda_0 = 3.68
\times10^{-7}$, a realistic rate corresponding to the geometric mean of
the estimated passenger rates across mutation types used in
\citet{Jones2008}. A geometric mean was used because of the
constraint on mutation type effects, that is, $\prod\alpha_m = 1$. Next,
$\alpha_m$'s were estimated from the data and samples effects
$\beta_k$'s were proportional to the numbers of mutations in the 24
samples in the pancreatic cancer data. Their products were set to 1
to satisfy the constraints. In summary, the sampling model for the
passenger genes in our simulation scenario is tailored to the data
and assumptions used in \citet{Jones2008}. The
mutation rates of a small set of randomly selected genes were inflated
to represent true drivers: 2\% of genes were set to have mutation rate
$10\lambda_0$, 1\% of genes were set to have mutation rate of
$30\lambda_0$ and 0.05\% of genes were set to have mutation rate
$200\lambda_0$. A 200-fold increase is realistic for the so-called
``mountain'' genes, while 10-fold and 30-fold increases correspond to the
``hills.'' The proportions of true drivers at different mutation rates
were chosen manually to make the overall distribution of observed
mutation counts close to that observed in the pancreatic cancer data.

%f1 #&#
\begin{figure}[b]

\includegraphics{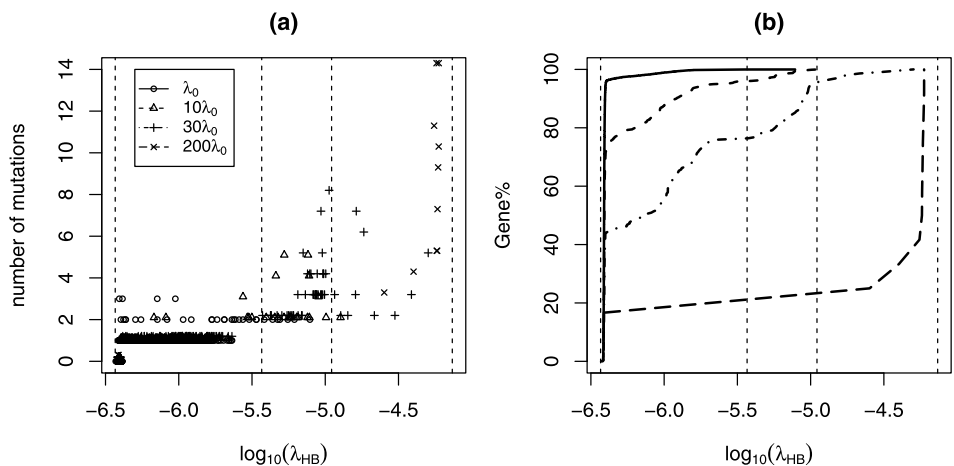}

\caption%[Estimated mutation rates from simulated data]
{\textup{(a)} Logarithm of estimated mutation rate ($\lambda_{\mathrm{HB}}$)
against the observed number of mutations. \textup{(b)}
Cumulative distribution of the logarithm of $\lambda_{\mathrm{HB}}$
with genes grouped by their true $\lambda_g$'s.
In \textup{(a)}, each point is one gene, and the Y axis levels
are slightly shifted to separate the groups. Vertical dashed lines
indicate true $\lambda_g$'s used in the simulation. The legend in
\textup{(a)} applies to both \textup{(a)} and \textup{(b)}.}
\label{figsimest}
\end{figure}

%f2 #&#
\begin{figure}[b]

\includegraphics{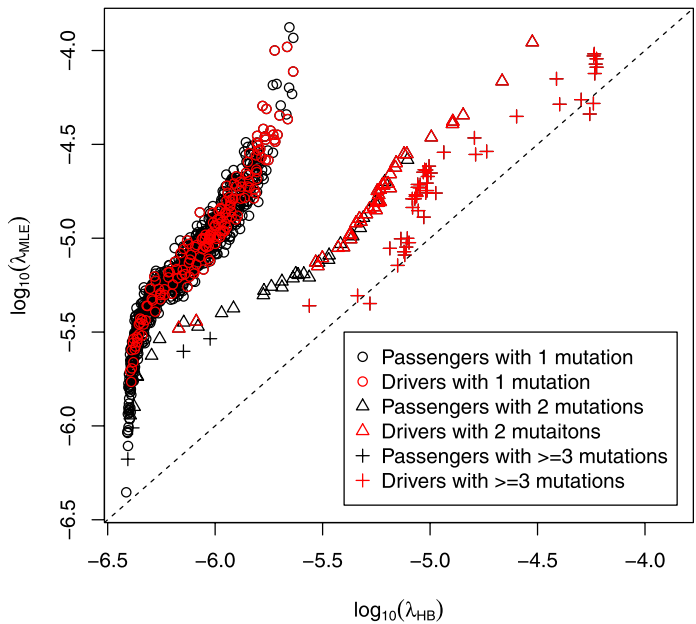}

\caption%[Hierarchical Bayesian estimates versus maximum likelihood
%estimates]
{Hierarchical Bayesian estimates versus maximum likelihood estimates
of mutation rates. Each point is a gene, labeled according to its
number of
mutations and colored according to whether it is a true
driver. Drivers are over-plotted or else drivers with a single
mutation would be invisible, given the large number of other
genes.}
\label{figsimhbmle}
\end{figure}

%s4.2 #&#
\subsection{Results}

Figure \ref{figsimest} shows results of mutation rate estimates from the
simulated data. Estimated $\lambda_g$'s of individual genes are shown
in Figure~\ref{figsimest}(a) against their observed mutation
counts. The average estimated mutation rate for genes with no
mutation is $3.83\times10^{-7}$, very close to the true
$\lambda_0=3.68\times10^{-7}$ used in the simulation, even though
$\lambda_0$ was not known to the estimation procedure. This suggests
that the model captures the underlying passenger mutation rate. Also,
for genes with no mutation, there is no
separation among genes with different true mutation rates, which is
expected since there is no information to distinguish them. Estimated
mutation rates generally increase as the number of mutations
increases, but there are also large differences in estimated rates
among genes with the same number of mutations, resulting from different
sizes and nucleotide compositions of those genes.

Figure \ref{figsimest}(b) shows the estimated $\lambda_g$'s by groups
defined by the true $\lambda_g$'s. Each line is the cumulative
distribution of the logarithms of the estimated $\lambda_g$'s for one
of the
groups. Even among genes with 200-fold increases over the passenger
rate, two genes are not distinguishable from passengers because they
did not have any mutations in the 24 simulated samples. This illustrates
the challenges of learning gene-specific mutation rates in this type
of study.

As an alternative approach to estimating mutation rates we considered
the maximum likelihood estimates (MLE) calculated for each gene
separately. We assumed the same Poisson model for the
MLE. For the calculation of the MLEs, the true
parameters $\alpha_m$'s and $\beta_k$'s were plugged in, a choice
that favors the MLEs. Figure \ref{figsimhbmle} compares
posterior means of $\lambda_g$'s obtained using our Hierarchical
Bayesian model to the MLEs. Only genes with at least one observed
mutation are shown. The
differences between the two approaches are striking. Ranking genes by
estimated rates, and proceeding down the list based on the Bayesian
estimates, one does not encounter a true passenger until position
39. On the other hand, the top two genes by MLE are both true
passengers and among the top 30 genes; only 22 are true drivers. The
behavior of the two approaches is most different for genes with a
single mutation, as expected. The hierarchical model has pulled these
strongly toward the overall genome mean, so that the genes with one
mutation rank below most of the genes with more than one mutation.
For genes with two mutations, the shrinkage is less pronounced, and for
genes with 3 or more mutations, the estimates are generally close,
with the exception of a small number of large genes who are pulled
strongly, and in a nonlinear pattern, toward smaller values.

The main difference between our hierarchical Bayesian approach and the
MLE is shrinkage. By using a mixing distribution representative of the
distribution of the genes' rates across the genome, the Bayesian
approach estimates each mutation rate using data from many other genes
with potentially similar rates. This underlying distribution is not
considered by the MLE approach.

%f3 #&#
\begin{figure}[b]

\includegraphics{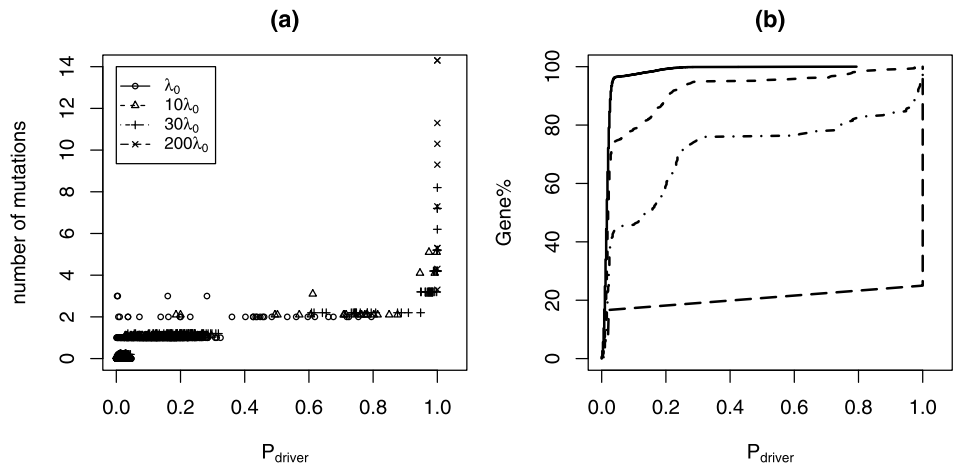}

\caption%[Estimated driver probabilities from simulated data]
{\textup{(a)} Estimated driver probability against the observed number of
mutations. \textup{(b)}~Cumulative distribution of the estimated
driver probabilities with genes grouped by their true
$\lambda_g$'s. In \textup{(a)}, each point is one gene, and the Y axis levels
are slightly shifted to separate the groups. The legend in \textup{(a)}
applies to both \textup{(a)} and \textup{(b)}.}
\label{figsimcla}
\end{figure}

Figure \ref{figsimcla} shows the posterior driver probabilities
from the same simulated data set. The true passenger mutation rate
used in the simulation was used as
$\lambda_0$ in the Bayesian model. Overall the results have
similar patterns compared to those of the estimated mutation
rates. Figure \ref{figsimcla}(a) shows estimated driver
probabilities of all genes against their observed mutation counts.
Genes with no mutation have estimated driver probabilities close to 0
regardless of their true mutation rates. As the number of mutations
increases, the estimated driver probabilities generally also
increase. Only a small number of genes have estimated probabilities
close to 1.
Figure \ref{figsimcla}(b) groups genes by their true mutation rates
to present the differences among the four groups. For genes with true
mutation rates equal to $200\lambda_0$, estimated driver probabilities
are large, except for the two genes with no observed mutation. A
substantial proportion of the genes with mutation rates equal to
$10\lambda_0$ and $30\lambda_0$ have estimated driver probabilities
much larger than 0.

The estimated proportion of driver genes, $\pi$, is 0.025 with
a 90\% credible interval $(0.017,0.041)$, while the true value used in
the simulation is 0.0305. We also used several different Beta
distributions as priors for $\pi$ and they all led to similar
posterior estimates. Using different values as $\lambda_0$ in the
model resulted in very different estimates of $\pi$. Doubling
$\lambda_0$ led to an estimated $\pi$ of 0.0065 with a 90\% credible
interval $(0.0047,0.0087)$, while reducing $\lambda_0$ by half led to an
estimated $\pi$ of 0.48 with a 90\% credible interval
$(0.37,0.59)$. These results show the dependence of the estimated $\pi$
on the input parameter $\lambda_0$.

\begin{figure}

\includegraphics{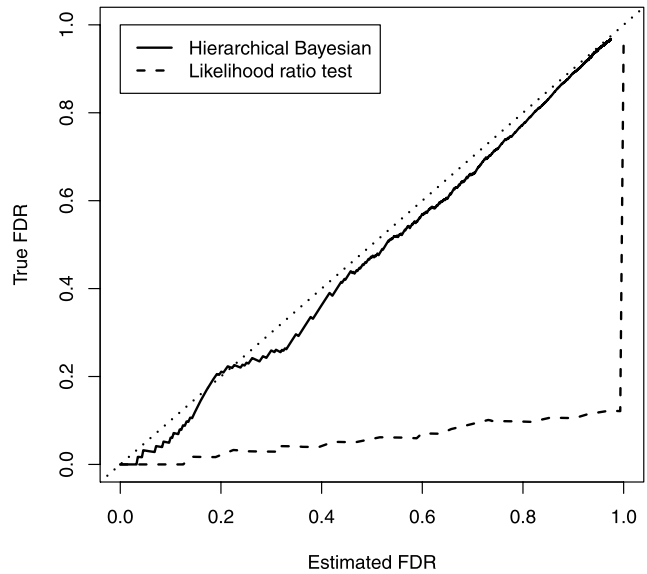}

\caption%[Estimated and true FDRs for hierarchical Bayesian
%and likelihood ratio test]
{True FDRs and estimated FDRs from hierarchical Bayesian estimates of driver
probabilities and likelihood ratio test $p$-values for all genes.}
\label{figsimfdr}
\end{figure}

We also used likelihood ratio tests (LRT) with Poisson
densities to analyze the simulated data. We used the true $\alpha_m$'s
and $\beta_k$'s for LRTs here. For gene $g$, under the null
hypothesis, $\lambda_g=\lambda_0$, the total number of mutations
$X_g=\sum_{m,k}(X_{\mathit{gmk}})$ follows a Poisson distribution with
parameter $\sum_{m,k}\alpha_m\beta_k T_{\mathit{gmk}}$. The $p$-value for
the likelihood ratio test can be calculated using the right-tail
probability of $X_g$ under the null hypothesis.
We then used the FDR controlling procedure from \citet{Benjamini1995}
to calculate estimated FDRs from LRT $p$-values. To compare the
results to those from our method, we also calculated estimated FDR
from Hierarchical Bayesian estimates of the driver
probabilities. True FDRs were calculated using the true driver
indicators used in the simulation. Figure \ref{figsimfdr} shows the
results from these two methods. The estimated FDRs from our
hierarchical Bayesian method are very close to the true FDRs, showed
by the closeness of the curve to the diagonal line. The estimated rates from
likelihood ratio tests are much smaller than the true rates,
suggesting that they are too conservative by as much as an order of
magnitude.

%f4 #&#

The main reason for the overestimation of FDR here is that the
controlling procedure assumes a uniform distribution of
$p$-values from true null tests. However, because the distribution of
mutation counts for each gene is Poisson and the mutation rate is very
small under the null hypothesis, the vast majority of true passenger
genes have mutation counts of 0. The resulting distribution of
$p$-values from true passenger genes is very different from a uniform
distribution. This shows that our method has substantially better
calibration and improved ability to estimate driver probabilities and
the overall proportion of driver genes compared to LRT coupled
with an FDR controlling procedure. This improvement is critical for
the appropriate interpretation of lists of candidate drivers and for
the efficient design of two-stage studies.

%s5 #&#
\section{Cancer mutation data analysis}
\label{sec5}
%s5.1 #&#
\subsection{Pancreatic cancer data}

\begin{figure}

\includegraphics{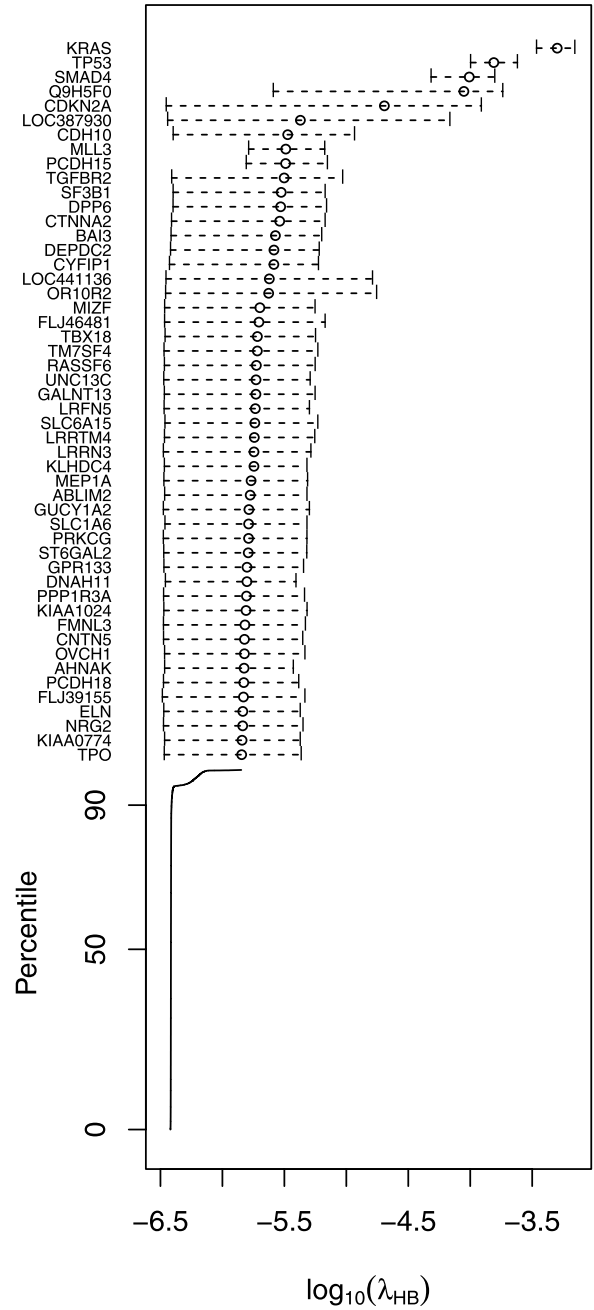}

\caption%[Estimate mutation rates from the pancreatic cancer data]
{Estimated mutation rates from the pancreatic cancer data.
Genes are ordered according to their estimated mutation rates
($\lambda_{\mathrm{HB}}$). The names and 90\% credible intervals of the top
50 genes are shown.}
\label{figrealest}
\end{figure}

Figure \ref{figrealest} shows the estimates of mutation rates with
the pancreatic cancer data. Genes are ordered by their estimated
mutation rates and the 50 genes with the largest estimated rates
are listed on the top. The mean of estimated mutation rates for
genes with no mutations is $3.93\times10^{-7}$, closer to the
``intermediate'' passenger mutation rate $\lambda_0=3.68 \times10^{-7}$
than the ``low'' rate $2.07\times10^{-7}$ and the ``high''
rate $5.30\times10^{-7}$ provided in \citet{Jones2008}. Among the top
50 genes, only a few have 90\% credible intervals completely above
the ``intermediate'' rate. Genes with small sizes, such as CDKN2A,
tend to have large credible intervals.

%f5 #&#

\begin{figure}

\includegraphics{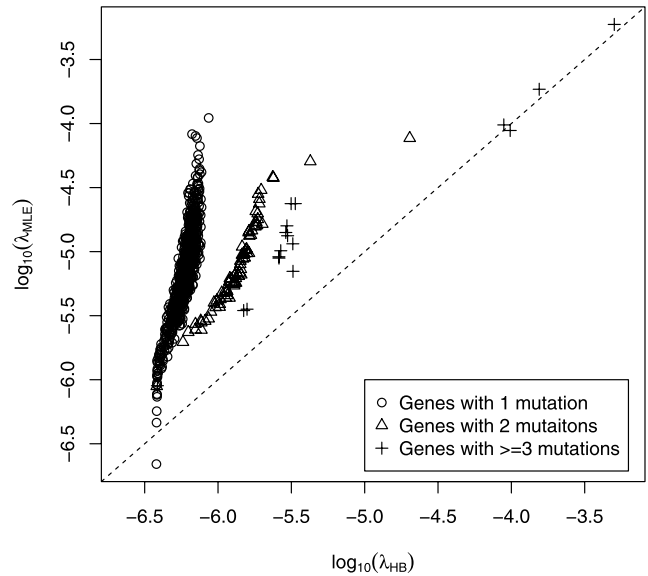}

\caption%[Hierarchical Bayesian estimates versus maximum likelihood
%estimates]
{Hierarchical Bayesian estimates versus maximum likelihood estimates
of mutation rates. Each point is a gene, labeled according to its
number of mutations.}
\label{figrealmle}
\end{figure}

We also calculated maximum likelihood estimates of the mutation
rates $\lambda_g$ for each gene with at least one observed
mutation. See supplementary material [\citet{Ding2012}] for the
details of MLE
calculation. The comparison between MLEs and hierarchical Bayesian
estimates is shown in Figure~\ref{figrealmle}. The overall shape
reproduces the pattern seen in the simulation study. The shrinkage
effect is evident for most genes with only 1 mutation, and it is
greater for small genes. For example, the gene OSTN, with only 300
bases sequenced, has a MLE of $5.7\times10^{-5}$, the 11th highest
rate, while the Bayesian estimate is only $7.3\times10^{-7}$, much
closer to the whole-genome average rate, and is ranked 117th.
On the other end, the gene PCDHGC4, with more than 52,000
bases sequenced, had a MLE of $2.2\times10^{-7}$ and a Bayesian
estimate of $3.8\times10^{-7}$, also closer to the genome
average. The MLEs and Bayesian estimates for genes with 3 or more
mutations are similar.

%f6 #&#

%f7 #&#
\begin{figure}

\includegraphics{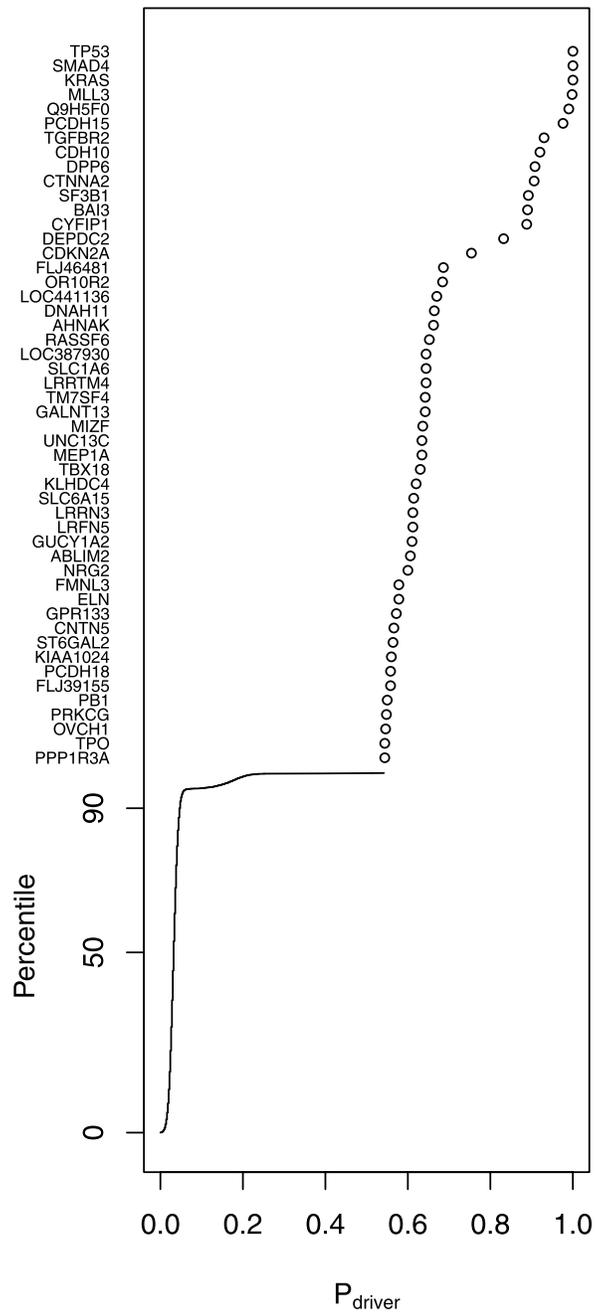}

\caption%[Estimated driver probabilities from the pancreatic cancer
%data]
{Estimated driver probabilities from the pancreatic cancer data.
Genes are ordered according to their estimated driver
probabilities ($P_{\mathrm{driver}}$). The names of the top 50 genes are shown.}
\label{figrealcla}
\end{figure}

Figure \ref{figrealcla} shows the estimated driver probabilities
using the ``intermediate'' rate from \citet{Jones2008} as the
passenger rate $\lambda_0$ in our model. Genes are
ordered by their estimated driver probabilities and the 50 genes with
the highest driver probabilities are listed on the top. The list of the
top 50 genes is very similar, though not identical, to that generated
by the estimated mutation rates. It is interesting to contrast the
inferences on genes CDKN2A and MLL3 with very different gene sizes.
CDKN2A is a small gene with 206 bases sequenced, so 2
mutations are enough to produce a large estimated mutation rate. CDKN2A
is ranked higher than MLL3, which is a much larger gene with 13,908
bases sequenced and 6 observed mutations. However, CDKN2As credible
interval is also much larger due to its small size. As a result, the
driver probability of MLL3 is close to one, while that of CDKN2A is
around 0.7, placing it far lower in the ranking.

The estimated proportion of driver genes, $\pi$, is
0.038 with 90\% credible interval $(0.018,0.066)$, corresponding to a
total number of drivers of 779 with credible interval $(381,1359)$. The
large credible interval and the numerous genes with driver probability
around 50\% highlight the challenge of classifying individual genes
using only 24 samples. However, the study provides strong evidence that
the total number of drivers in pancreatic cancer is large.

Changing input passenger mutation rate has a large effect on the estimates
of driver probabilities and on the overall proportion of drivers.
Using the ``high'' passenger mutation rate resulted in an estimated
$\pi=0.0041$ with 90\% credible interval $(0.0016,0.0080)$, while using the
``low'' rate resulted in an estimated $\pi=0.28$ with 90\% credible
interval $(0.21,0.37)$. These rates are likely to be conservative upper
and lower bounds. While the posterior driver probabilities are
affected by the choice of passenger mutation rate $\lambda_0$, their
relative orders are much more robust. For example, using the
``high'' rate produced a list of top 50 genes which share 38 genes with
the top 50 list using the ``low'' rate. Also, even when using a
conservative upper bound on the passenger mutation rate, the expected
number of drivers is close to~100.

\begin{figure}

\includegraphics{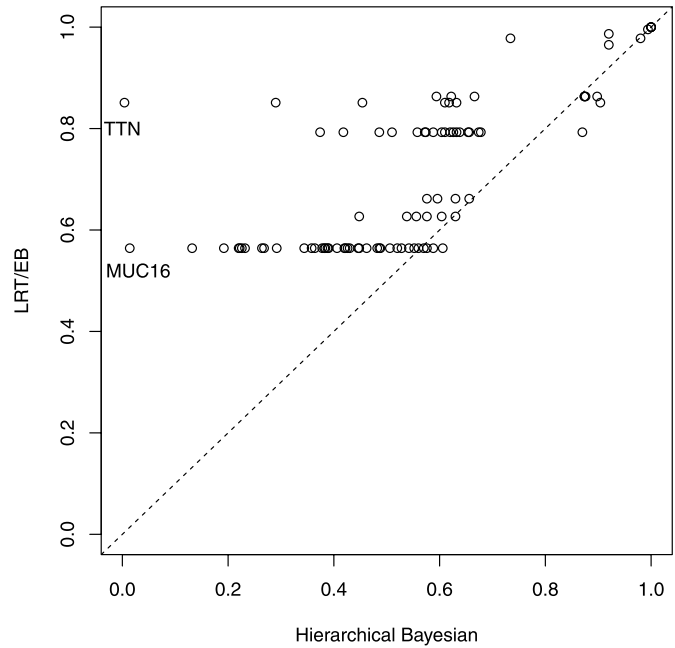}

\caption%[Comparison between estimated driver probabilities]
{Comparison between estimated driver probabilities from
the hierarchical Bayesian method (HB) and from the
likelihood ratio test/empirical Bayes method (LRT/EB) in
Jones et~al. (\citeyear{Jones2008}).}
\label{figrealhbeb}
\end{figure}

The original paper analyzing the pancreas cancer data
[\citet{Jones2008}] used an empirical Bayes local FDR method of
\citet{Efron2002}, constructed using the likelihood ratio test proposed
in \citet{Getz2007}. Figure \ref{figrealhbeb} compares driver
probabilities estimated using the hierarchical Bayesian model in this
paper to the probabilities estimated in \citet{Jones2008}. Only genes
with 2 or more mutations are plotted in the figure. This is done so
the list of genes is roughly the same as the list of genes in the
table S7 in \citet{Jones2008}. Note that the table in
\citet{Jones2008} also used amplification and deletion data, which
are not used in the comparison here. The estimates from these two methods
are positively correlated. For most genes shown in the figure, estimated
probabilities using our method are lower than those estimated using the
empirical Bayes approach. The granularity of the estimates from
\citet{Jones2008} arises from the conservative steps taken to overcome
statistical and numerical difficulties of estimating a null distribution
when event rates are low, and from monotonization of the FDR
estimates. Our Bayesian approach, through shrinkage, smoothness and
other features, provides a higher resolution. It also provides a
different ranking. To illustrate, the genes TTN and MUC16 are
highlighted in Figure \ref{figrealhbeb} on the left.
TTN has 6 mutations but also has more than 100K bases sequenced, the
most in this data set. This causes a greater discounting in the
hierarchical Bayes approach than the MLE-based empirical Bayes
approach. This is consistent with the shrinkage
pattern observed in Figure \ref{figsimcla}.
The other example is MUC16, which has 2
mutations and 40K bases sequenced, the third most in this data set.
Another factor that may account for some of the differences in
ranking is the consideration of sample effects, not used in
\citet{Jones2008}.\eject

\begin{figure}

\includegraphics{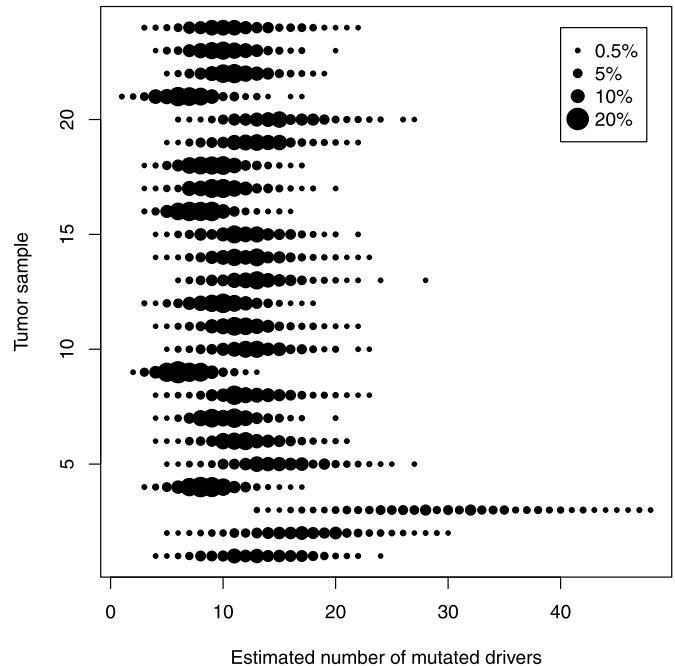}

\caption%[Estimated number of mutated drivers]
{Posterior distribution of the estimated number of mutated drivers
in each tumor sample.}
\label{figrealmutdriver}
\end{figure}

As another summarization of the hierarchical Bayesian results,
Figure \ref{figrealmutdriver} shows the posterior distribution of
the estimated number of mutated drivers in each tumor sample. All
samples except one have at least two mutated drivers among all
posterior simulations. The remaining one has less
than 1\% posterior probability of having only one
mutated driver. Most samples harbor five or more mutated drivers with
high probabilities; the average number is 12.

%f8 #&#

%f9 #&#

%s5.2 #&#
\subsection{Breast cancer data}

The breast cancer genome project [\citet{Wood2007}] is presented
here to
emphasize the flexibility of the Bayesian approach in dealing with
two-stage designs. The sample size was smaller
than that of the pancreatic cancer data. Because of that, the results
have more variability. The estimated mutation rates $\lambda_g$ range
from $8.6\times10^{-7}$ to $1.35\times10^{-4}$. The average mutation
rate for genes with no mutation is $1.23\times10^{-6}$, much higher
than the corresponding rate in the pancreatic cancer data. This rate is
again closest to the intermediate, or ``SNP-based,'' passenger mutation
rate among the three estimation methods in \citet{Wood2007}.
The estimated overall driver
proportion $\pi$ varies for different passenger mutation rates used in
the model. Using ``External,'' ``SNP-based'' and ``NS/S-based''
passenger rate estimates resulted in $\pi$ estimates of 53\%, 12\% and
0.02\%, respectively.

%s6 #&#
\section{Discussion}
\label{sec6}

We developed a hierarchical Bayesian methodology to estimate
gene-specific mutation rates and driver probabilities as well as the
proportion of drivers among sequenced genes from somatic
mutation data in cancer.

To distinguish driver genes from passenger genes solely based on
marginal mutation rates, somewhat strong assumptions are needed. The
first is that all passenger genes have the same mutation rate.
Biologically, mutation rates can vary across different regions of
genome [\citet{Wolfe1989}] from factors such as DNA replication timing
[\citet{Wolfe1989,Stamatoyannopoulos2009}] and chromatin structure
[\citet{Prendergast2007,Schuster2012}]. With the sample sizes available
in the data sets analyzed in this paper, it is difficult to consider
variation in passenger rates explicitly, though ongoing sequencing
effort may allow a deeper exploration of this issue in the near future.

Another key assumption is that mutations in different
genes occur independently. Because of this assumption, we can estimate
a gene's driver probability using\vadjust{\goodbreak} its marginal mutation
rate. In practice, it is likely
that mutations in one gene can lead to growth advantage or
disadvantage depending on whether certain mutations in some other
genes exist or not, especially if these genes are in the same
biological pathway. While modeling of such interactions is possible
for selected pathways [\citet{Boca2010,Ciriello2012}],
estimation of even pairwise dependencies at the gene level across the
entire genome remains challenging.

These assumptions represent a reasonable compromise between the
limitations of available sequencing data and the need to prioritize
candidate driver genes for further research in a model-based way. They
were commonly made in other cancer somatic mutation studies
[e.g., \citet{Jones2008,CancerGenomeAtlasResearchNetwork2008}]. With
the development of new sequencing technologies and the
increasing amount of cancer sequencing data, new methodologies will be
needed, likely with a more flexible set of assumptions.

Our model also assumes that each sample is homogeneous such that if a
mutation occurs in a gene in one sample, it occurs in all cells from
that sample. This assumption realistically models the data generated
by Sanger sequencing with strict quality control, where only
mutations shared by the majority of cells are identified. In
reality, cancer samples are often heterogeneous: the same
sample can distinct subpopulation of cancer cells at different stages
of evolution or even following from different evolutionary paths. So
a certain mutation may only present in a proportion of cells. Such
information can be obtained using deep sequencing technologies
available now [\citet{Walter2012}]. To analyze
such data, an additional layer could be incorporated into the
hierarchical Bayesian models to account for the heterogeneity of cells
in a sample. A challenge in modeling this information will arise from
the fact that mutations in different genes can have different levels
of heterogeneity.

We designed two models, one for estimating gene-specific mutation
rates and one for estimating gene-specific driver probabilities and
the overall proportion of drivers. Both
achieved similar results in terms of separating groups of genes with
different true mutation rates in the simulation study and ordering the
top candidate driver genes in the pancreatic and breast cancer genomes
data. While estimating driver probabilities provides a more direct way
to answer the question of distinguishing drivers from passengers, the
model does depend on the assumption that there is a single underlying
passenger mutation rate common to all passenger genes and requires
this rate as an input parameter.

So far, most analyses of somatic mutations rely on external estimates of
the mutation rates for passenger genes, obtained, for example, from
sequencing data from noncoding regions or rates of silent mutations
[\citet{Wood2007}]. This input has a large effect on the estimated
proportion of driver genes and\vadjust{\goodbreak}  the overall magnitude of the driver
probabilities. However, the order of top candidates is not affected
substantially either in simulated or real data. We thus recommend the
use of estimated mutation rates for ranking, selection and prediction,
as the model for this estimation does not require any assumption on
the passenger mutation rate, nor does it need an estimate of this
rate. In either model, Bayesian modeling allows us to use these
external estimates, when available, for specifying the prior
distribution.

Both models in this paper use a Dirichlet process on the unknown
distribution of the gene-specific mutation rates across the genome. This
assumption can be substituted with other types of distributions, including
parametric ones. For example, we considered a log-normal distribution
for mutation rate estimation and a mixture prior with point mass at
$\lambda_0$ and a log-normal distribution truncated at $\lambda_0$ for
driver probability estimation. When we applied these two choices to
the simulated data, model fit was not as satisfactory as that of the
Dirichlet process (see supplementary material [\citet{Ding2012}]
for details),
likely because there were a few genes
with very high mutation rates (the mountains) together with a much
larger set of genes with moderately increased mutation rates (the
hills). The Log-normal distribution does not fit this situation well,
nor would most of the commonly used parametric distributions, especially
if unimodal and controlled by a small number of parameters. Thus, we
strongly recommend the use of a flexible distribution, which can be
estimated reliably even in relatively small studies, if the number of
genes is large.

Results provided here are but examples of many summarizations one can
produce using the MCMC output. For example, for each gene one can
easily compute the predictive probability of observing a mutation in a
hypothetical new tumor sample or new study. Another useful approach is to
examine gene sets or pathways. The model output can be used to compute
the probability that a chosen pathway is altered by one or more
driver mutations in each of the patients, as suggested in \citet{Boca2010}.

In an important paper \citet{Greenman2006} provided
likelihood-based testing approaches for distinguishing drivers from
passenger mutations. An interesting aspect of their work is the
modeling of both the mutation process and the selection pressure on
the tumor. They also considered the significance of selection toward
missense, nonsense and splice site mutations, and proposed tests
assessing variation in selection between functional domains. A
combination of the approach considered here with the features
introduced by \citet{Greenman2006}, while well beyond the scope of this
article, could potentially be very useful.

Our methodology provides estimates of the total number of driver genes.
The early cancer genome project highlighted the importance of ``hills,''
or genes that are drivers in a relatively small proportion of
tumors. Increasing independent evidence is accumulating to support
the importance of the hills. Hills are numerous and easy to miss in
small studies, which suggests that many more undiscovered\vadjust{\goodbreak}  hills may
exist. Our model attempts a quantification of the size of this
population based on mutation rates alone. This quantification is
difficult, whence the large credible intervals, and sensitive to
assumptions on passenger rates. Nonetheless, our method leads to the
prediction that the population is large, very likely in the hundreds,
and possibly in the thousands.

In conclusion, our models produce posterior inferences on all relevant
parameters, using data generated from single, multi-stage and multiple
studies, potentially sequencing different sets of genes. We expect that
these tools will be helpful in both assessing the evidence provided by
existing data and in planning further experiments to confirm the genes'
role in cancer development.

%s7 #&#
\section{Software}

An R package is freely available at
\href{http://bcb.dfci.harvard.edu/\%7Egp/software/CancerMutationMCMC/}{http://bcb.dfci.harvard.}
\href{http://bcb.dfci.harvard.edu/\%7Egp/software/CancerMutationMCMC/}{edu/\%7Egp/software/CancerMutationMCMC/}.

\begin{supplement}[id=suppA]
\stitle{Supplementary methods and results}
\slink[doi]{10.1214/12-AOAS604SUPP} %[doi,text={...}] - jei reikia
%suskaldyti doi
\sdatatype{.pdf}
\sfilename{aoas604\_supp.pdf}
\sdescription{Additional technical details and simulation results.}
\end{supplement}

% imsref loaded by lrinkeviciute, 2013-01-09 10:04:52
% imsref loaded by lrinkeviciute, 2013-01-09 10:20:31

\printaddresses

\end{document}